\documentclass[letterpaper]{article} % DO NOT CHANGE THIS
\usepackage{aaai25}  % DO NOT CHANGE THIS
\usepackage{times}  % DO NOT CHANGE THIS
\usepackage{helvet}  % DO NOT CHANGE THIS
\usepackage{courier}  % DO NOT CHANGE THIS
\usepackage[hyphens]{url}  % DO NOT CHANGE THIS
\usepackage{graphicx} % DO NOT CHANGE THIS
\usepackage{natbib}  % DO NOT CHANGE THIS AND DO NOT ADD ANY OPTIONS TO IT
\usepackage{caption} % DO NOT CHANGE THIS AND DO NOT ADD ANY OPTIONS TO IT
\usepackage[]{algorithm2e}

\usepackage{amsfonts}
\usepackage{amsmath}
\usepackage{amssymb}
\usepackage{amsfonts}
\usepackage{multirow}
\usepackage{multicol}
\usepackage{cleveref}
\usepackage{subcaption}
\usepackage{xfrac}
\usepackage{booktabs}
\usepackage{bm}

\usepackage{tikz}
\usetikzlibrary{shapes, arrows, positioning}
\title{Accurate Computation of the Logarithm of Modified Bessel Functions on GPUs\footnote{Accepted at ICS 2024}}

\author {
    Andreas Plesner\textsuperscript{\rm 1}\footnote{Corresponding author \texttt{aplesner@ethz.ch}},
    Hans Henrik Brandenborg Sørensen\textsuperscript{\rm 2},
    Søren Hauberg\textsuperscript{\rm 2}
}
\affiliations {
    \textsuperscript{\rm 1}ETH Zurich\\
    \textsuperscript{\rm 2}Technical University of Denmark
}

\newtheorem{corollary}{Corollary}

\begin{document}

\maketitle

\begin{abstract}

Bessel functions are critical in scientific computing for applications such as machine learning, protein structure modeling, and robotics. 
However, currently, available routines lack precision or fail for certain input ranges, such as when the order $v$ is large, and GPU-specific implementations are limited.
We address the precision limitations of current numerical implementations while dramatically improving the runtime.
We propose two novel algorithms for computing the logarithm of modified Bessel functions of the first and second kinds by computing intermediate values on a logarithmic scale.
Our algorithms are robust and never have issues with underflows or overflows while having relative errors on the order of machine precision, even for inputs where existing libraries fail.
In C++/CUDA, our algorithms have median and maximum speedups of 45x and 6150x for GPU and 17x and 3403x for CPU, respectively, over the ranges of inputs and third-party libraries tested. 
Compared to SciPy, the algorithms have median and maximum speedups of 77x and 300x for GPU and 35x and 98x for CPU, respectively, over the tested inputs.

The ability to robustly compute a solution and the low relative errors allow us to fit von Mises-Fisher, vMF, distributions to high-dimensional neural network features. 
This is, e.g., relevant for uncertainty quantification in metric learning. 
We obtain image feature data by processing CIFAR10 training images with the convolutional layers of a pre-trained ResNet50. 
We successfully fit vMF distributions to 2048-, 8192-, and 32768-dimensional image feature data using our algorithms. 
Our approach provides fast and accurate results while existing implementations in SciPy and mpmath fail to fit successfully.

Our approach is readily implementable on GPUs, and we provide a fast open-source implementation alongside this paper.%\looseness=-1
\end{abstract}

% \keywords{Bessel Functions, von Mises-Fisher, GPU, CUDA, Robust Computation}

\section{Introduction}
Bessel functions, an integral part of scientific computing, find widespread applications in various fields, including machine learning, protein structure modeling, and robotics \citet{oh2020radial, boomsma2008generative, hadi:rss:2021}. These functions, in their standard and modified forms, arise as solutions to Bessel differential equations \citet{watson1922treatise, abramowitz1972handbook}.
\begin{figure}[ht!]
    \centering
    \begin{subfigure}{\columnwidth}
        \includegraphics[width=\linewidth]{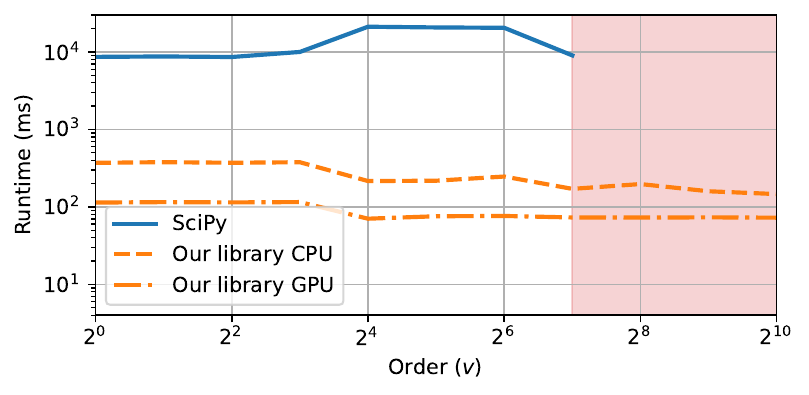}
        \caption{
            Runtime comparison between our library and SciPy. The runtime for each value of $v\{2^0,2^1,...,2^{10}\}$ is measured by sampling 20M values for $x$ in the interval $[1, 100]$. 
            SciPy underflows in the red region ($v \geq 128$). 
            For all values of $v$, the runtime of our library is one to two orders of magnitude faster than that of SciPy. 
            % We also run the test for the modified Bessel function of the second kind, and when $v\leq16$, our library is slower than SciPy. 
            % However, across the two modified Bessel functions, our library sees a median speedup of, respectively, 35x and 77x for CPU and GPU, and a maximum speedup of, respectively, 98x and 300x for CPU and GPU.    
        }
        \vspace{.1cm} % Add some vertical space between the figures
        % \label{fig: sub1}
    \end{subfigure}
    \begin{subfigure}{\columnwidth}
        \includegraphics[width=\linewidth]{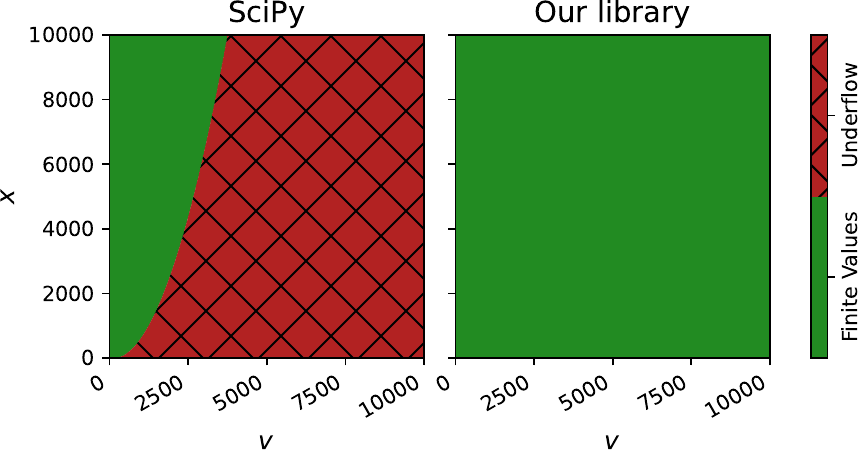}
        \caption{Stability comparison for computing $\log I_v(x)$: SciPy versus our library. The left panel shows SciPy's underflow regions (in red), indicating limited stability. In contrast, the right panel demonstrates that our library consistently returns finite values (in green), reflecting its high robustness. The color scheme highlights the computational reliability of our library over the tested domain.}
        \vspace{-0.1cm}
        % \label{fig: sub2}
    \end{subfigure}
    \caption{Comparison of the robustness and runtime of Scipy's scaled implementation and our library for calculating the logarithm of the modified Bessel function of the first kind.} 
    \label{fig: scipy comparison}
    % \Description{Comparison between the runtime of our library (CPU and GPU versions) and SciPy. The three curves are roughly constant for all values of the order. The curve for SciPy stops before the others as SciPy fails when the order is 256 or larger.}
    \vspace{-0.2cm}
\end{figure}
Their relevance extends to physical phenomena such as vibrations, hydrodynamics, and heat transfer \citet{bowman2012introduction, korenev2002bessel}.

This paper focuses on the modified Bessel functions of the first and second kinds, denoted, respectively, $I_v(x)$ and $K_v(x)$, where $v$ represents the order and $x$ the argument. These functions will be formally defined in \Cref{sec: theory}.

In statistics, the modified Bessel functions are central to distributions such as the Mises-Fisher distribution and the K-distribution \citet{watson1983statistics, mardia2000directional, kotz2004continuous}. Despite their widespread use, there remain significant challenges with their numerical computation; particularly in terms of robustness when dealing with large order values and over different input ranges, as they easily underflow or overflow. Although early work has been done on algorithms that optimize accuracy and reduce computational time, they still have significant problems \citet{bremer2019algorithm, lin2016improved, biagetti2016discrete, tumakov2019faster}.\looseness=-1

In Bayesian deep learning, the limitations of current methods for computing modified Bessel functions have been noted, with existing techniques often lacking precision or prone to overflow or underflow making them unusable \citet{oh2020radial}. 
To circumvent these problems, \citet{oh2020radial} resort to bounds on the ratio $\frac{I_{v/2-1}(x)}{I_{v/2}(x)}$ when $v$ takes large values, e.g.\@ $v$ in the thousands or tens of thousands. \Cref{fig: scipy comparison} demonstrate these limitations using SciPy's implementation, which struggles beyond $v=128$ and exhibits longer runtimes than our library. % From the figure we can see that SciPy can only calculate the functions up to $v=128$. Additionally, the runtime is much longer for SciPy than for our library.

To address the accuracy limitations of current implementations, we focus on computing the logarithm of modified Bessel functions efficiently and effectively. Some methods for computing Bessel functions focus on scaling the Bessel functions to avoid numerical problems such as underflow or overflow. In these cases, the Bessel function $B(x)$ is scaled by a function $S_B(x)$, which for $I_v(x)$ and $K_v(x)$ are given by \Cref{eq: scale function Iv,eq: scale function Kv}. In this way, the scaling functions compensate for exponential increases and decreases in the functions. SciPy uses these scaling functions to extend the range of input in which it can compute the Bessel functions \citet{2020SciPy-NMeth}.

\begin{align}
    S_I(x) &= \exp(-\left|x\right|),\label{eq: scale function Iv}\\
    S_K(x) &= \exp(\left|x\right|).\label{eq: scale function Kv}
\end{align}
Instead of scaling the functions with exponentials, we propose to compute the logarithms of the modified Bessel functions, $\log I_v(x)$ and $\log K_v(x)$, by computing intermediate values on a logarithmic scale to ensure the numerical stability of the results.

In addition to the lack of robustness, the existing implementations are not fast and are CPU-bound. Therefore, we also develop a GPU version, since CUDA only has implementations for $v=0$ and $v=1$, and applications need $\log I_v(x)$ for any $v>0$.

This paper is part of a project that seeks to create a special function library that can run on GPUs. This avoids GPU-to-CPU and CPU-to-GPU memory transfers when performing machine learning with special functions, as these memory transfers are expensive \citet{gregg2011data}. The functions should be precise and have runtimes comparable to existing solutions when using a GPU.\footnote{The code can be found at \url{https://lab.compute.dtu.dk/cusf/cusf}} \looseness=-1

% \textbf{The paper is structured as follows.}~~ The next section reviews existing libraries to compute Bessel functions and their limitations. \Cref{sec: theory} covers the theory of Bessel functions and how we can compute them on a logarithmic scale. \Cref{sec:algorithms} develop fast and accurate numerical algorithms from the theoretical results, and \Cref{sec: data for numerical experiments} presents the experimental setup used for validation. Lastly, \Cref{sec: numerical results} experimentally show that our approach significantly improves, robustness, accuracy, and running time over existing approaches, and showcases an application in machine learning. %\looseness=-1

\section{Related Work}

\paragraph{Applications of Bessel Functions.} The modified Bessel function of the first kind appears in the probability density function of the von Mises-Fisher distribution. This models data on the $p\!-\!1$ unit hypersphere, $S^{p-1}=\{ \mathbf{x}\in\mathbb{R}^p ~|~ \|x\|_2=1 \}$, and is a cornerstone of directional statistics \citet{mardia2000directional}; the von Mises-Fisher distribution in $\mathbb{R}^p$ uses $I_v(x)$ for $v=\sfrac{p}{2}$ and $v=\sfrac{p}{2}-1$ (\Cref{sec: metric learning}). We review a few applications next.\looseness=-1

\citet{hadi:rss:2021} create a generative model with the von Mises-Fisher (vMF) distribution to capture end-effector orientations in a robot control setting.

The vMF distribution is also used in metric learning, which aims for similar data to be near, while dissimilar ones should be far away \citet{musgrave2020metric}. 
\citet{warburg:metric:2023} propose a Bayesian method to capture the inherent uncertainty of the model predictions. Their pipeline embeds images into a neural network feature space, which is normalized to unit norm. Using a Laplace approximation to capture weight uncertainty, they approximate the predicted feature distribution with a vMF distribution, which requires high-order Bessel functions. 

\citet{boomsma2008generative} use the vMF distribution to model local protein structures. \citet{banerjee2005clustering} use vMF to cluster high-dimensional data and give methods to fit the vMF distributions without having to evaluate the modified Bessel function. However, the approximation has errors around $1 \%$ even for low-dimensional data, $p\in[10,100]$. 

\citet{oh2020radial} use the vMF distribution for Bayesian uncertainty quantification and observe the inherent instability in modern libraries. Therefore, they create their own approximation to estimate the ratio between the modified Bessel function of the first kind of orders $\sfrac{p}{2}$ and $\sfrac{p}{2-1}$. This is required when estimating the parameters of the vMF distribution using the statistical estimation method presented by \citet{sra2012short}. This shows that whenever the modified Bessel functions were encountered, the researchers had to come up with approximations because they did not have a numerically stable method for computing the functions, which is the core contribution of our work.\looseness=-1

\paragraph{Calculating Bessel Functions.}
Many software libraries that handle special functions incorporate algorithms to compute Bessel functions, often based on the approach described by \citet{amos1986algorithm}. An example is the Boost C++ library, which implements the unscaled Bessel functions. The documentation of the Boost library contains tests of the precision of their implementations; they report the relative errors in units of epsilon (machine precision) using Mathematica for the reference solutions. The tests are done with double precision, but no input range is given. The relative errors they report are around machine precision for doubles ($\approx2\cdot10^{-16}$), but we experimentally found (\Cref{sec: numerical results}) that the relative errors can be several orders of magnitude larger. % when using Mathematica for the reference values.

\paragraph{Calculating Logarithm of Bessel Functions.}
Research has already been conducted on computing the logarithm of Bessel functions. For example, \citet{rothwell2006} presents algorithms for both the Bessel functions of the first and second kind, using recursion in areas where the conventional approach introduced by \citet{amos1986algorithm} faces numerical challenges. However, it is important to recognize that the recursion method means that the runtimes for computing the logarithms grow linearly with the order $v$. In addition, these methods involve the use of complex numbers in the computation process because the Bessel functions can be negative. The high runtime when $v$ is large would negatively affect the performance on GPUs as we wish to process large arrays of numbers and not just singular values. Additionally, complex numbers increase the memory load and register usage, thus limiting the maximum number of concurrent kernels. Similarly, \citet{cuingnet2023computation} propose an approach for the logarithm of the modified Bessel function of the second kind, but the work does not provide code or comparative information, making any comparison difficult.

\citet{bremer2019algorithm} provides algorithms to evaluate the logarithm of the Bessel function of the first and second kind, with errors ranging from $10^{-12}$ to $10^{-9}$ and a runtime of approximately 3 to 4 seconds for 10M input values. Errors are estimated using Mathematica to provide reference solutions. Our implementation manages to do much better for the logarithm of the modified Bessel of the first and second kind; we get errors around machine precision ($10^{-16}$) while the runtime in most cases is 1 to 2 orders of magnitude better.

% \citet{xue2018recursive} present methods for computing the logarithm of ratios of modified Bessel functions. 

\section{Theory}\label{sec: theory}

This section will go into the theory necessary to develop the algorithms and introduce asymptotic expressions. \citet{watson1922treatise} gives a thorough explanation of the Bessel functions.

\textbf{Defining Bessel Functions.}~~
The Bessel functions are defined as the solutions, $y(z)$, to Bessel's differential equation shown in \Cref{eq: bessel differential equation}, where $z\in\mathbb{Z}$ is the argument and $v$ is the order \citet{watson1922treatise}. Meanwhile, the modified Bessel functions are the solutions to Bessel's modified differential equation seen in \Cref{eq: modified bessel differential equation}, where $x\in\mathbb{R}$ is now the argument \cite[p. 374]{abramowitz1972handbook}.
\begin{align}
    z^2\frac{\mathrm{d}^2 y}{\mathrm{d}z^2} + z\frac{\mathrm{d} y}{\mathrm{d}z} + (z^2 - v^2)y&=0,\label{eq: bessel differential equation} \\
    x^2\frac{\mathrm{d}^2 y}{\mathrm{d}x^2} + x\frac{\mathrm{d} y}{\mathrm{d}x} - (x^2 + v^2)y&=0\label{eq: modified bessel differential equation}.
\end{align}

The Bessel functions of the first and second kind are the solutions to \Cref{eq: bessel differential equation}, where $y(z)$ is, respectively, finite or divergent for $z=0$. The solutions are respectively denoted as $J_v(z)$ and $Y_v(z)$.
The modified Bessel functions of the first and second kind, $I_v(x)$ and $K_v(x)$, are given by restricting $J_v(z)$ and $Y_v(z)$ to purely imaginary inputs $z$; for example, $I_v(x)=i^{-v}J_v(ix),x\in\mathbb{R}$ \citet{DLMF}.

\textbf{Logarithmic computations.}~~
When computing logarithms for Bessel functions, a certain transformation is essential that involves logarithms of sums. It is the ``logarithm-of-a-sum'' trick for a sequence of $N$ positive numbers, providing a formula to efficiently compute the logarithm of the sum. This technique improves numerical accuracy, for instance, by focusing on the logarithms of terms rather than the terms themselves. The transformation is given as
\begin{align}
    \log\sum_{k=0}^N a_k &= \log a_i + \log\sum_{k=0}^N \frac{a_k}{a_i} \nonumber\\
   &= \log a_i + \log\sum_{k=0}^N \exp(\log a_k - \log a_i).\label{eq: log sum trick}
\end{align}
For the sake of numerical accuracy, the optimal $a_i$ to select is the largest term of the series, i.e. $i = \arg\max_k a_k$.
This prevents the exponentiation operations from overflowing.

\subsection{Modified Bessel function of the first kind}

First, we will consider the modified Bessel function of the first kind, $I_v(x)$, for the inputs $v\geq 0$ and $x \geq 0$. This function produces positive outputs, so the logarithm is well-defined. An infinite series can be used to evaluate the function for all inputs \cite[p. 375 Eq. 9.6.10]{abramowitz1972handbook}. However, it is advantageous to use some asymptotic expressions for various ranges of input values to speed up the computations and improve the numerical accuracy. The same techniques and some of the expressions used for $I_v(x)$ can be applied with minor modifications to the modified Bessel function of the second kind, $K_v(x)$. $I_v(x)$ is given by the infinite series below \citet{DLMF}.
\begin{align}
    I_v(x) = \left(\frac{x}{2}\right)^v \sum_{k=0}^\infty \frac{\left(\frac{x^2}{4}\right)^k}{k!\Gamma(k+v+1)}.\label{eq: Iv infinite series}
\end{align}

To compute $\log I_v(x)$, we truncate the series after a finite number of terms. For this, we have the following corollary. 
\begin{corollary}
    \label{co: Iv converges}
    The series in \Cref{eq: Iv infinite series} converges absolutely for all inputs.
\end{corollary}
\noindent
This follows from the fact that the factorial function grows much faster than the exponential function.

\Cref{co: Iv converges} implies that, in practice, the series can be truncated after $N$ terms.
\begin{align*}
    I_v(x) \approx \left(\frac{x}{2}\right)^v \sum_{k=0}^N \frac{\left(\frac{x^2}{4}\right)^k}{k!\Gamma(k+v+1)}.
\end{align*}

\textbf{Recurrence relation.}~~
Given the sum presented above, we can simplify the calculation of the terms by introducing a recurrence relation. 
In the series \Cref{eq: Iv infinite series}, the terms are given by \Cref{eq: ak terms} and can be calculated recursively, \Cref{eq: ak recurrence base,eq: ak recurrence}. This is relevant for efficient implementations, as the previous terms can be used to quickly calculate the next term.
\begin{align}
    a_k &= \frac{\left(\frac{x^2}{4}\right)^k}{k!\Gamma(k+v+1)},\label{eq: ak terms}\\
    a_0 &= \frac{1}{\Gamma(v+1)}, \label{eq: ak recurrence base}\\
    a_{k+1}&= a_k \frac{x^2}{4(k+1)(k+v+1)}\label{eq: ak recurrence}.
\end{align}
Taking the logarithm of $I_v(x)$ and applying the  ``logarithm of a sum'' trick gives
\begin{align}
    \log I_v(x) &\approx v\log\frac{x}{2} + \log a_i \nonumber\\
    &+ \log\sum_{k=0}^{N} \exp\left(\log a_k - \log a_i \right). \label{eq: log Iv}
\end{align}
where $a_i$ is the largest of the terms $\log a_k$;
\begin{align*}
    i =\arg\max_k& \log a_k\\
    = \arg\max_k& \bigg\{ k\log\left(\frac{x^2}{4}\right) -\sum_{j=1}^k \log j - \log\Gamma(k+v+1) \bigg\}.
\end{align*}

Thus, to evaluate $\log I_v(x)$, we only need the logarithm of the terms $a_k$.
Due to the multiplication in the recurrence relation for $a_k$, \Cref{eq: ak recurrence}, the logarithm of the terms can be efficiently evaluated without the risk of underflow or overflow using a similar recursion,
\begin{align}
    \log a_0 &= -\log\Gamma(v+1), \\\nonumber
    \log a_{k} &= \log a_{k-1} + 2\log x \nonumber\\
    &- \log 4 - \log k - \log(k+v). \label{eq: log ak recurrence}
\end{align}

With these equations, a computational routine can be developed to compute $\log I_v(x)$. Given the input parameters $v$ and $x$, the procedure involves the following steps:
\begin{enumerate}
    \item Initialize the value of $N$ to ensure sufficient accuracy.
    \item Compute \Cref{eq: log ak recurrence} for the $a_k$ terms.
    \item Calculate $\log I_v(x)$ using \Cref{eq: log Iv}.
\end{enumerate}

This leaves an unanswered question. What value $N$ should be used to accurately calculate $\log I_v(x)$? 

Using the recurrence relation for $\log a_k$ as given in \Cref{eq: log ak recurrence}, the value of $\log a_k$ changes by, 
\begin{align*}
    \log a_{k+1} - \log a_k = &2\log x - \log 4 \\
    &- \log(k+1) - \log(k+v+1).
\end{align*}

Thus, the value of $\log a_k$ increases $\forall k < K$, where $K$ is the unique value that satisfies $$2\log x - \log 4 = \log(K) + \log(K+v),$$ and decreases $\forall k > K$.

Solving the equation for $K$ gives \Cref{eq: K} with the approximation holding when $x \gg v$, $x \ll v$, and $x \approx v$. Thus, the location of the peak term is approximately linear in $x$. 
\begin{align}
    K = \frac{-v \pm \sqrt{x^2+v^2}}{2} \approx \frac{x}{2} \label{eq: K}.
\end{align}

The computations are done in finite precision, so when doing the summation in \Cref{eq: log sum trick}, we get no influence on the summation $\forall k:\log a_k - \log a_K < \log\epsilon$, where $\epsilon$ is the machine precision. The terms grow (almost) exponentially until this point, followed by a superexponential ($c/(k!)$) decrease, so many terms would not affect the summation.\footnote{One could sort the terms in the summation to avoid this; however, as the terms rapidly decrease far away from $K$ the impact is negligible, and it involves doing an expensive sorting operation. We later show that the numerical precision remains good without this sorting.} 

Therefore, it is not necessary to compute all the terms, as only the terms above machine precision in \Cref{eq: log sum trick} affect the output. Terms less than machine precision, relative to the maximum value, are ignored. 
From empirical experiments, we observe that the number of relevant terms grows as $9.2\sqrt{x}$ when $x\gg v$ or $x \approx v$. However, if $x$ is large, then $\sqrt{x}$ is still large, so it is still necessary to evaluate many terms. Therefore, we will use asymptotic expressions when the inputs are large.

\textbf{Assymptotic expressions.}~~
We find asymptotic expressions that hold when $v$ or $x$ is large to avoid evaluating the summation presented above for many terms.
We found two asymptotic expressions that are fast to evaluate and give accurate results when the series definition of $\log I_v(x)$ is lacking. Numerical tests against MATLAB's Bessel functions are used to determine in which regions the expressions work well; we will focus on this in \Cref{sec:algorithms}.

The first expression is denoted as the ``$\mu_K$ expression'' and is given below for $\log I_v(x)$ with $\mu=4v^2$ \citet{DLMF}.
\begin{align}
    \log I_v(x) &\sim x - \frac{1}{2}\log(2\pi x) \nonumber\\
    &+ \log\left| 1 + \sum_{k=1}^\infty (-1)^k \frac{\Pi_{j=1}^k(\mu - (2j-1)^2)}{k!(8x)^k}\right|.\label{eq: log Iv mu k}
    % \mu &= 4v^2.\nonumber
\end{align}
This expression works well for large arguments $x$ provided the order $v$ is small. Numerical tests have shown that it is not necessary to evaluate more than about 20 terms in the infinite series. The main reason for using 20 terms instead of, say, 5 terms is that the expression works for slightly smaller inputs. However, this effect plateaus after 20 terms. The $K$ in $\mu_K$ indicates that the first $K$ terms are used to calculate the result.
There is an absolute value around the infinite series, as numerical errors can cause it to be negative. In addition, as seen earlier with the $\log a_k$ recursion, the terms in the series can also be calculated recursively to improve the runtime.

The second expression is denoted as the ``$U_K$ expression'' and is given below for $\log I_v(x)$ \citet{DLMF}.
\begin{align}
    \log I_v(x) &\sim -\frac{1}{2}\log(2 \pi v) + v \eta - \frac{1}{4}\log(1+x'^2) \nonumber\\
    &+ \log\left| 1 + \sum_{k=1}^\infty \frac{u_k(t)}{v^k}\right|,\label{eq: log Iv u k}\\
    x'=\frac{x}{v}, t &= \frac{1}{\sqrt{1+x'^2}}, \eta = \sqrt{1+x'^2} + \log \frac{x'}{1+\sqrt{1+x'^2}}.\nonumber
\end{align}
The infinite series is cut off after a few terms and the $K$ is again used to indicate how many terms are kept. This method can only be used when $v\neq 0$ due to the definition of $x'$, and it does not work well if $0<v\ll 1$.
The functions $u_k(t)$ are polynomials given by the recursive expression \citet{DLMF}.
\begin{align}
    u_0(t) =& 1,\label{eq: u0}\\
    u_{k+1}(t) =& \frac{t^2 - t^4}{2} \frac{\mathrm{d}}{\mathrm{d}t} u_k(t) 
    + \frac{1}{8}\int_0^t (1 - 5x^2)u_k(t) \mathrm{d}x.\label{eq: uk}
\end{align}
We have only been able to find the polynomials $u_k(t)$ for $k=1,...,6$ in the literature \cite[10.41(ii)]{DLMF}. Using Mathematica, we computed $u_k(t)$ for $k=7,...,13$. 
Like in \Cref{eq: log Iv mu k}, the infinite series in \Cref{eq: log Iv u k} contains an absolute value sign, which is only needed to limit numerical issues.

Based on the input values, we then have three available methods, the sum definition from \Cref{eq: Iv infinite series}, the $\mu_k$ expression from \Cref{eq: log Iv mu k}, and lastly the $U_K$ expression from \Cref{eq: log Iv u k} to calculate $\log I_v(x)$. In \Cref{sec:algorithms} we will determine the input ranges for each method.

\subsection{Modified Bessel function of the second kind}
This subsection focuses on the elements necessary to calculate the logarithm of the modified Bessel function of the second kind, $\log K_v(x)$, for input argument $x$ and order $v$. Starting with the asymptotic expressions, $\mu_K$ and $U_K$ can still be used with minor changes, as indicated below.

\textbf{Asymptotic expressions.}~~
The ``$\mu_K$ expression'' for $\log K_v(x)$ is given below with $\mu=4v^2$ \citet{DLMF}.
\begin{align}
    \log K_v(x) &\sim \frac{1}{2}(\log\pi - \log(2x)) - x \nonumber\\
    &+ \log\left| 1 + \sum_{k=1}^\infty \frac{\Pi_{j=1}^k(\mu - (2j-1)^2)}{k!(8x)^k}\right|.\label{eq: log Kv mu k}
    % \mu &= 4v^2. \nonumber %\\
\end{align}
Here, only some of the first coefficients and the alternating sign in the infinite series have changed. Therefore, the considerations for the $\mu_K$ expression of $\log I_v(x)$ still hold. Specifically, the same number of terms and input ranges works well.

The ``$U_K$ expression'' for $\log K_v(x)$ is similar to before \citet{DLMF}.
\begin{align}
    \log K_v(x) &\sim \frac{1}{2}(\log\pi - \log(2v)) - v\eta - \frac{1}{4}\log(1 + x'^2) \nonumber\\
    &+ \log\left| 1 + \sum_{k=1}^\infty (-1)^k \frac{u_k(t)}{v^k}\right|,\label{eq: log Kv u k}\\
    x'=\frac{x}{v}, t &= \frac{1}{\sqrt{1+x'^2}}, \eta = \sqrt{1+x'^2} + \log \frac{x'}{1+\sqrt{1+x'^2}},\nonumber
\end{align}
with $u_k(t)$ being as in \Cref{eq: u0,eq: uk}.

\textbf{Integral definition.}~~
The asymptotic expressions only hold for large inputs, so we need an expression for small values. For small inputs, we can use the integral expression in \Cref{eq: log Kv integral} \cite[p. 241 Eqs. (26)-(27)]{rothwell2006}. 
\begin{align}
    \log K_v(x) &= \frac{1}{2}\log\pi - \log\Gamma(v+\frac{1}{2}) -v\log(2x)-x \nonumber \\
    &+\log\int_0^1 \beta\exp(-u^\beta)(2x+u^\beta)^{v-\frac{1}{2}}u^{n-1}\nonumber\\
    &\hspace{1cm}+\exp{\frac{-1}{u}}u^{-2v-1}(2xu+1)^{v-\frac{1}{2}} \mathrm{d}u, \label{eq: log Kv integral} \\
    \beta&= \frac{2n}{2v+1}, n=8. \nonumber
\end{align}
 To simplify expressions based on the integral, we define the functions $f(u)$, $g(u)$, and $h(u)$ as given below.
\begin{align*}
    f(u) &= \beta\exp(-u^\beta)(2x+u^\beta)^{v-\frac{1}{2}}u^{n-1},\\
    &+\exp{\frac{-1}{u}}u^{-2v-1}(2xu+1)^{v-\frac{1}{2}} = g(u) + h(u), \\
    g(u) &= \beta\exp(-u^\beta)(2x+u^\beta)^{v-\frac{1}{2}}u^{n-1}, \\
    h(u) &= \exp{\frac{-1}{u}}u^{-2v-1}(2xu+1)^{v-\frac{1}{2}}.
\end{align*}
We need to evaluate the integral numerically, which we will do using Simpson's 1/3 composite rule \citet{solomon2015numerical}.
\begin{align*}
    \int_0^1 f(u) \mathrm{d}u &\approx\frac{4}{6N}\sum_{k=1}^{\frac{N}{2}} f((2k-1)h) + \frac{2}{6N}\sum_{k=1}^{\frac{N}{2}-1} f((2k)h) + \\
    &+\frac{f(0)+f(1)}{6N},\\
    h&=\frac{1}{N}.
\end{align*}
The $N$ is constant, and numerical tests have shown that $N=600$ gives acceptable results balancing runtime and accuracy for all inputs tested. 
To simplify the sums we define the weights $w_k$ given by:
\begin{align*}
    w_k = \begin{cases}
        4, k\text{ odd}\\
        2, k\text{ even}
    \end{cases}.
\end{align*}
Combining the weights with the definition of $g$ and $h$, and applying the ``logarithm-of-a-sum'' trick, the logarithm of the integral term in \Cref{eq: log Kv integral} can be rewritten as follows. Note that $f(0)=0$.
\begin{align*}
    &\log\int_0^1 f(u) \mathrm{d}u \approx - \log(6N)+ \log M \\
    &\quad+ \log\left( \exp{(\log G - \log M)} + \exp{(\log H - \log M)} \right), \\
    &\hspace{1.1cm}\log M = \max(\log G,\log H),\\
    &\hspace{1.1cm}\log G = \log\left( \sum_{k=1}^{N-1} w_k g(kh) + g(1) \right), \\
    &\hspace{1.1cm}\log H = \log\left( \sum_{k=1}^{N-1} w_k h(kh) + h(1) \right).
\end{align*}
The integral can then be computed quickly from $\log G$ and $\log H$. We once again apply the ``logarithm-of-a-sum'' trick; however, we precompute the maximum values before evaluating the sum. Otherwise, we must store all terms in memory and loop through the sums twice. To simplify, we focus solely on $\max g(u)$ and $\max h(u)$ and ignore the weights $w_k$. 

\textbf{Finding heuristics for maximums.}~~
The maximum value of a function, here $g(u)$ and $h(u)$, is either where the derivative is zero or the boundaries of the input range. Based on the integral \Cref{eq: log Kv integral} we have $u\in[0,1]$. Therefore, the maximum can also be at $u=0$ and $u=1$. 

We start by considering $g(u)$, and we apply the logarithm as this will not change the location of the maximum. Furthermore, we take the derivative with respect to $u$.
\begin{align*}
    \log g(u) =& \log\beta -u^\beta + \left(v - \frac{1}{2}\right) \log (2x + u^\beta) \\
    &+ (n-1)\log u, \\
    \frac{\mathrm{d}}{\mathrm{d}u}\log g(u) =& -\beta u^{\beta-1} + \left(v - \frac{1}{2}\right) \frac{\beta u^{\beta-1}}{2x + u^\beta} + \frac{n-1}{u}.
\end{align*}
Setting the derivative to zero and solving for $u$ gives %the following.
\begin{align*}
    1 = \frac{v - \frac{1}{2}}{2x + u^\beta} + \frac{n-1}{\beta u^\beta}.
\end{align*}
Setting $x=u^\beta$ and solving for $x$ gives a quadratic equation with solutions given below in \Cref{eq: x}.
\begin{align}
    x &= \frac{-2x\beta+v\beta-\frac{\beta}{2}+n-1}{2\beta} \nonumber\\
    &\pm\frac{\sqrt{(-2x\beta+v\beta-\frac{\beta}{2}+n-1)^2+8\beta x (n-1) }}{2\beta}\label{eq: x}.
\end{align}
If there exists a solution $x\in\mathbb{R}$ to the above equation such that $u^*=x^{\frac{1}{\beta}}\in[0,1]$ then $u^*$ is the argument that maximizes $g$. However, it is difficult to determine for which input ranges solutions exist. If $u^*$ exists, numerical experiments have shown $u^*\in[0.95,1]$, with the maximum only slightly larger than $g(1)$. Furthermore, in most cases there are no solutions $u^*\in[0,1]$, and the maximum is at $u=1$. So, the heuristic $u^* = 1$ is valid in most cases, and if it is false, then $\max_u g(u) - g(1)$ is small. Thus, there are no numerical problems when using this heuristic. 

We will then consider $\log h(u)$ and its derivative. In addition, we will set the derivative to zero and solve for $u$.
\begin{align*}
    \log h(u) &= \frac{-1}{u} (-2v-1)\log u \left(v-\frac{1}{2}\right) \log(2xu+1), \\
    \frac{\mathrm{d}}{\mathrm{d}u}\log h(u) &= \frac{1}{u^2} - \frac{2v+1}{n} + \frac{2x\left(v-\frac{1}{2}\right)}{2xu+1},\\
    2v+1 &= \frac{1}{u} + \frac{v \frac{1}{2}}{1 + \frac{1}{2xu}}.
\end{align*}
The last expression can be rewritten as a quadratic equation with solutions:
\begin{align*}
    u &= \frac{-2(v-x) -1 \pm \sqrt{(2v-2x+1)^2 + 4(2vx+3x)}}{4vx + 6x}.
\end{align*}
The part under the square root is non-negative for all inputs $x$ and $v$. Thus, we see that a solution always exists, and numerical tests have shown this to be the maximum in all cases. However, the following heuristic has been shown to approximate the solution $u^*=\arg\max h(u)$ very well.
\begin{align*}
    u^* = \begin{cases}
        \frac{1}{2}, & v < 2 \\
        \frac{1}{2v},& v \geq 2
    \end{cases}.
\end{align*}
Thus, we can get a good approximation of the maximums of $g(u)$ and $h(u)$ without evaluating complex expressions.\looseness=-1

We can now calculate $\log K_v(x)$ using these three available methods, the integral definition \eqref{eq: log Kv integral}, the $\mu_k$ expression \eqref{eq: log Kv mu k}, and the $U_K$ expression \eqref{eq: log Kv u k}. In \Cref{sec:algorithms}, we determine suitable input ranges for each method.

\section{Algorithms}\label{sec:algorithms}
The expressions presented above must be merged to give algorithms to calculate $\log I_v(x)$ and $\log K_v(x)$. To do this, we need to determine which expression to use for a given input. \looseness=-1

TEST: \ref{sec:algorithms}

We conducted tests of the runtimes in MATLAB 2020b and found that the $\mu_K$ expression is the fastest to compute for all $K\leq 20$. The second fastest is the $U_K$ expression for all $K\leq 13$. Finally, the infinite series and integral expressions, for, respectively, $\log I_v(x)$ and $\log K_v(x)$, are the slowest. 
% The order of a method is the number $K$, that is, $\mu_{20}$ is the $\mu$ method of order 20. The $U_K$ methods are computationally heavy, so using the lowest order possible is beneficial. 
% The result is \Cref{algo: log Iv or log Kv} which can be used to calculate $\log I_v(x)$ for $v\geq 0$ and $\log K_v(x)$ for all inputs as the same branching can be used to calculate both.

\subsection{Choosing the right expressions}
We have different approximations of $\log I_v(x)$ and $\log K_v(x)$ and use numerical experiments to determine which is more accurate for different input ranges. The tests show that the same input ranges are valid for the $\mu_K$ and $U_K$ expressions when calculating both $\log I_v(x)$ and $\log K_v(x)$. The input ranges can be seen in \Cref{tab: input ranges}.

\begin{table}
    \centering
    \begin{tabular}{c|p{6cm}}
        Expression & Input ranges\\\hline
        $\mu_3$     & $((x > 1400) \wedge (v<3.05)) \vee (([0.6229\log{x} -3.2318] > \log{v}) \wedge (v > 3.1))$ \\\hline
        $\mu_{20}$  & $( (x > 30) \wedge (v<15.3919) ) \vee ( ([0.5113\log{x} + 0.7939] > \log{v}) \wedge (x > 59.6925) )$ \\\hline
        $U_4$       & $(x > 274.2377 \wedge v > 0.3) \vee (v > 163.6993)$ \\\hline
        $U_6$       & $(x > 84.4153 \wedge v > 0.46) \vee (v > 56.9971)$ \\\hline
        $U_9$       & $(x > 35.9074 \wedge v > 0.6) \vee (v > 20.1534)$ \\\hline
        $U_{13}$    & $(x > 19.6931 \wedge v > 0.7) \vee (v > 12.6964)$ \\\hline
    \end{tabular}
    \caption{Input ranges where the asymptotic expressions for the modified Bessel functions of the first and second kind are applicable. Ordered top-to-bottom from fastest to slowest.}
    \label{tab: input ranges}
    \vspace{-0.1cm}
\end{table}

The order of the expressions in the table gives the order of priority for the expressions with the series and integral definitions as the fallback cases. Thus, if for an input $(v,x)$ the expression $\mu_{20}$ is valid, then it will be used regardless of whether any of the $U_K$ expressions can be used. Additionally, to simplify the code, we do not use all $\mu_K$ or $U_K$ expressions for $K=3,...,20$ and $K=4,...,13$, respectively. While the $\mu_4$ expression is faster than the $\mu_{20}$ expression, the difference is not very large, so to simplify the code, we chose a select few expressions based on numerical tests. The main runtime sinks are currently the sum and integral expressions, while the other expressions are much faster.

\subsection{Pseudo code}
Using the expressions presented previously and the input ranges for which they are valid, we can write a routine to calculate $\log I_v(x)$ and $\log K_v(x)$ given an input $(v,x)$. The pseudocode can be found in Algorithm~\ref{algo: cpu log Iv or log Kv}.

\begin{algorithm}[ht]
 \KwData{$x,v$}
 \KwResult{$\log I_v(x)$ or $\log K_v(x)$}
 initialization\;
    \uIf{$((x > 1400) \wedge (v<3.05)) \vee (([0.6229\log{x} -3.2318] > \log{v}) \wedge (v > 3.1))$}{
        Use $\mu_3$ expression\;
    }
    \uElseIf{$( (x > 30) \wedge (v<15.3919) ) \vee ( ([0.5113\log{x} + 0.7939] > \log{v}) \wedge (x > 59.6925) )$}{
        Use $\mu_{20}$ expression\;
    }
    \uElseIf{$(x > 274.2377 \wedge v > 0.3) \vee (v > 163.6993)$}{
        Use $U_4$ expression\;
    }
    \uElseIf{$(x > 84.4153 \wedge v > 0.46) \vee (v > 56.9971)$}{
        Use $U_{6}$ expression\;
    }
    \uElseIf{$(x > 35.9074 \wedge v > 0.6) \vee (v > 20.1534)$}{
        Use $U_{9}$ expression\;
    }
    \uElseIf{$(x > 19.6931 \wedge v > 0.7) \vee (v > 12.6964)$}{
        Use $U_{13}$ expression\;
    }
    \Else{
        Use series definition (for $\log I_v(x)$) or integral definition (for $\log K_v(x)$)\;
    }
 \caption{Compute $\log I_v(x),v\geq 0$ or $\log K_v(x),v\in\mathbb{R}$ using the equations given in the theory section when running on a CPU. When running on a GPU the branches for the $\mu_3, U_4, U_6, U_9$ expressions are removed to reduce warp divergence.}
 \label{algo: cpu log Iv or log Kv}
\end{algorithm}

Using these algorithms, we can then compare them to the solutions available in other software packages by directly comparing the runtimes and accuracies, but also by testing them on a real-world use case. 

\subsection{GPU specific optimizations}
The CPU code uses naive parallelization by parallelizing the loop over the elements using OpenMP. We make some modifications to the CPU code to optimize the code for GPUs. The code is written for CUDA where threads are collected in blocks of threads; 256 threads in our case gave good results. The threads in a block are collected in warps of 32 threads, where each warp is used for the Single Instruction Multiple Threads, SIMT, execution model. When naively parallelized on GPUs, the $i$'th thread computes $\log I_v(x)$ or $\log K_v(x)$ for the $i$'th input value $(v_i,x_i)$ as for CPUs. However, the threads in a warp should work on the same expression. For example, if the $i$'th and $i+1$'th elements are computed using the $\mu_{20}$ and $U_{9}$ expressions, respectively, and they run on the same warp, this will cause warp divergence. To avoid this and to balance the load for an entire block of threads, we sort the input elements based on which expression is used for each element. This means that if there are $k$ elements computed using the $\mu_{20}$ expression, then $\left\lfloor \frac{k}{256} \right\rfloor$ blocks will work entirely on the $\mu_{20}$ expression, resulting in high utilization for these blocks. Measurements of the runtime will include the sorting operations, which account for $\approx33\%$ of the runtime, however without the sorting the code is overall 3 to 4 times slower. Furthermore, as noted in Algorithm~\ref{algo: cpu log Iv or log Kv}, the GPU version omits some of the expressions. This is done as an additional measure to improve utilization and thus performance. The faster expressions that cover smaller input ranges have been removed, which has been found to improve the overall runtime for GPUs.

\section{Data and experimental setup for numerical experiments}\label{sec: data for numerical experiments}

\begin{table}
    \centering
    \begin{tabular}{c|c|c}
        Region                  & Function      & Number of Reference Solutions \\\hline
        \multirow{3}{*}{Small}  & $\log I_v(x)$ & 999,341                       \\
                                & $\log K_v(x)$ & 1,000,000                     \\
                                & $\log I_0(x)$ & 10,000,000                    \\\hline
        \multirow{3}{*}{Large}  & $\log I_v(x)$ & 605                           \\
                                & $\log K_v(x)$ & 39                            \\
                                & $\log I_0(x)$ & 10,000                        \\
    \end{tabular}
    \caption{Number of reference solutions for testing the precision when calculating $\log I_v(x)$, $\log K_v(x)$ and $\log I_0(x)$.}
    \label{tab: reference solutions fractional zero order}
    \vspace{-0.1cm}
\end{table}

We benchmark our library against the corresponding functions for $\log I_v(x)$ and $\log K_v(x)$ in the standard library (std), the GNU Scientific Library (GSL), and the Boost library \citet{CppStandard, GSL, Boost}. For GSL we can use their scaled functions, which reduce the risk of numerical underflows and overflows. We take the logarithm and add or remove the argument to undo the scaling functions shown in \Cref{eq: scale function Iv,eq: scale function Kv}, while for all others, we take the logarithm of the function outputs. 
Moreover, GSL, Boost, and the CUDA Math Library (CUDA in tables) contain functions specifically for orders $v=0$ and $v=1$. Therefore, we also compare our library's general $\log I_v(x)$ function against these special-purpose functions.

\subsection{Test Regions}
We evaluated the performance of $\log I_v(x)$ and $\log K_v(x)$ in two distinct regions of $(v,x)$. The two regions that were considered for testing are the \emph{Small region} ($(v,x)\in[0,150]\times[0,150]$) and the \emph{Large region} ($(v,x)\in[150, 10000]\times[150, 10000]$ for $\log I_v(x)$, and $(v,x)\in[150, 4000]\times[150, 4000]$ for $\log K_v(x)$\footnote{For large values the $K_v(x)$ function in Mathematica did not terminate so we limited the Large region}).
When testing the special case $v=0$, we use the Small and Large regions for the input argument $x$. The Small region is chosen so that third-party libraries, specifically Boost, could compute values for $\log I_v(x)$ without having underflow or overflow errors or returning Not a Number (NaN). The large region was limited by the range of inputs for which it was feasible to compute reference values using Mathematica.

\subsection{Test Points Sampling}
For performance evaluation of $\log I_v(x)$ and $\log K_v(x)$, we sampled 10M points uniformly in the specified regions, while for $v=0$, we uniformly sample 100M points in both regions. \looseness=-1

For precision testing, we uniformly sample 1M points in the Small region and 1K points in the Large region for the fractional order. We sample 1M points in the Small region and 10K points in the Large region for the special case $v=0$.\looseness=-1

\subsection{Reference Solutions}
To establish reference solutions for precision tests, we utilized Mathematica 13.3 to perform accurate calculations for each sampled point and stored up to 16 decimal points to ensure that reference solutions are precise up to machine errors for doubles. For precision testing, any points that were incorrectly evaluated in Mathematica were filtered out. Such inaccuracies were determined by Mathematica not providing a numeric answer but simply stating "Indeterminate". However, it should be noted that even when Mathematica provides a numeric output, it may not be accurate in some cases; particularly when evaluating values where $v\approx 100$ and $x\approx 0.1$. In such cases, Mathematica raises a warning indicating that there is a loss of precision. To obtain more accurate results, we suggest using Wolfram$|$Alpha. However, due to its computational heaviness and time-consuming nature, we only used Wolfram$|$Alpha for a few dozen input points, which will be discussed later.

The reference solutions provide a baseline for evaluating the accuracy of the implemented functions on both CPU and GPU.\looseness=-1

\section{Numerical results}\label{sec: numerical results}

\begin{table*}[ht]
\centering
\footnotesize
\begin{tabular}{c|c|c|c|c|c|c|c}
\toprule
\multirow{2}{*}{Function} & \multirow{2}{*}{Region} & \multirow{2}{*}{Metric} & \multicolumn{5}{c}{Library} \\
                        &                         &                           & std & GSL & Boost & CUSF (CPU) & CUSF (GPU) \\\midrule
\multirow{6}{*}{$\log I_v(x)$}  & \multirow{3}{*}{Small}  & Robustness & $100 \%$ & $99.98 \%$ & $\bm{100 \%}$ & $100 \%$ & $100 \%$ \\
                        & & Median     & $4.04 \times 10^{-16}$ & $1.34 \times 10^{-16}$ & $0.0$ & $2.12 \times 10^{-16}$ & $2.08 \times 10^{-16}$ \\
                        & & Maximum & $2.77 \times 10^{-6}$ & $2.03 \times 10^{-7}$ & $4.10 \times 10^{-8}$ & $8.30 \times 10^{-4}$ & $8.30 \times 10^{-4}$ \\
\cmidrule{2-8}
                        & \multirow{3}{*}{Large}  & Robustness & $0.50 \%$ & $44.13 \%$ & $1.98 \%$ & $\bm{100 \%}$ & $\bm{100 \%}$ \\
                        & & Median     & $1.20 \times 10^{-16}$ & $0.00$ & $0.00$ & $2.40 \times 10^{-16}$ & $2.28 \times 10^{-16}$ \\
                        & & Maximum & $1.20 \times 10^{-5}$ & $1.46 \times 10^{-15}$ & $0.00$ & $2.98 \times 10^{-13}$ & $2.98 \times 10^{-13}$ \\
\midrule
\multirow{6}{*}{$\log K_v(x)$}  & \multirow{3}{*}{Small}  & Robustness & $99.91 \%$ & $99.91 \%$ & $99.91 \%$ & $\bm{100 \%}$ & $\bm{100 \%}$ \\
                        & & Median     & $0.00$ & $0.00$ & $0.00$ & $1.61 \times 10^{-16}$ & $1.61 \times 10^{-16}$ \\
                        & & Maximum & $1.38 \times 10^{-11}$ & $2.29 \times 10^{-11}$ & $1.23 \times 10^{-11}$ & $6.50 \times 10^{-9}$ & $6.50 \times 10^{-9}$ \\
\cmidrule{2-8}
                        & \multirow{3}{*}{Large}  & Robustness & $100 \%$ & $82.05 \%$ & $\bm{100 \%}$ & $\bm{100 \%}$ & $\bm{100 \%}$ \\
                        & & Median     & $1.19 \times 10^{-16}$ & $1.32 \times 10^{-16}$ & $1.31 \times 10^{-16}$ & $2.40 \times 10^{-16}$ & $2.40 \times 10^{-16}$ \\
                        & & Maximum & $1.41 \times 10^{-5}$ & $5.02 \times 10^{-8}$ & $ 5.02 \times 10^{-8} $ & $5.02 \times 10^{-8}$ & $5.02 \times 10^{-8}$ \\
\bottomrule
\end{tabular}
\caption{Precision metrics for different libraries in Small and Large regions when calculating $\log I_v(x)$ and $\log K_v(x)$. The errors are the absolute relative errors compared to the reference solutions calculated using Mathematica. The robustness value in bold indicates the library with the highest robustness value, where the maximum relative error is used to break ties. for a given function and region. When computing $\log I_v(x)$ for the Small region, notice that the boost library has a much lower maximum error than the other libraries when using Mathematica as the reference solution. }
\label{tab: combined_precision}
\vspace{-5mm}
\end{table*}

This section looks at the results of experiments performed using the data presented in the previous section to test different libraries that compute $\log I_v(x)$ and $\log K_v(x)$.
The tests were run on an NVIDIA RTX 2080 TI GPU and an Intel Xeon Silver 4208 CPU using all 16 cores. We tested later on newer hardware (NVIDIA A100 80GB and AMD EPYC 7742 CPU using all 64 cores), and the relative differences in runtimes between libraries and the precision was unchanged. Tests are performed using double precision floating point numbers, which reduces the throughput of arithmetic instructions on the GPU by a factor of 32 compared to using single precision floating point numbers \cite[5.4.1. Arithmetic Instructions]{cuda}. We are more concerned with numerical accuracy than runtime, so the potential throughput improvement from 32 times more arithmetic instructions does not offset the reduced accuracy.

\subsection{Precision}
We split results for the CPU and GPU functions in our library and denote the computing device explicitly.
The precision is measured as the absolute value of the relative error to the reference solution. We compute precision as the median and maximum statistics, excluding Not-A-Number (NaN) and $\pm$infinity values. In cases where all values are NaN or $\pm\infty$, the median and the maximum are denoted as Not Available (N/A). \looseness=-1

We define \emph{robustness} as the fraction of test points for which the methods were able to produce an answer that was not NaN or $\pm\infty$. \emph{Robustness} is the key metric when comparing libraries, and we break ties with the maximum statistic.

\begin{figure}[th]
    \centering
    \includegraphics[width=\linewidth]{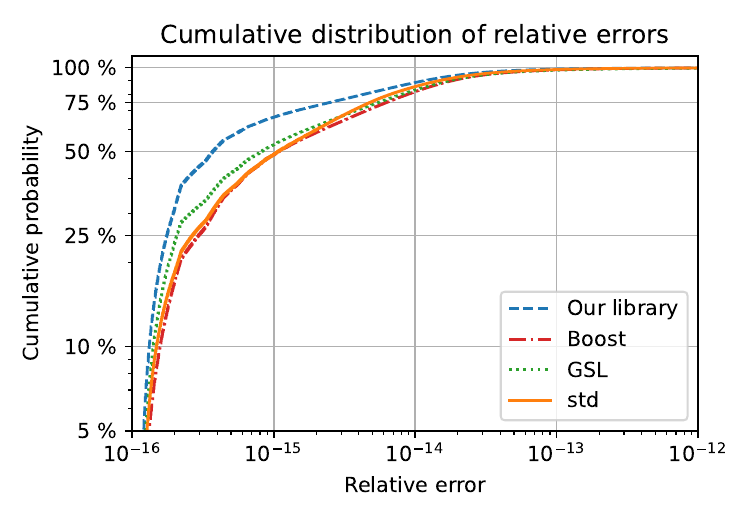}
    \vspace{-8mm}
    \caption{Cumulative distribution of relative errors for our and the three third-party libraries when computing $\log I_v(x)$ for the Small region. The plot illustrates the comparative performance in terms of numerical precision. Our library has a steeper curve, indicating a higher proportion of outputs with minimal relative errors, demonstrating our library's higher precision in evaluating the logarithm of the modified Bessel function of the first kind over the entire range tested.\looseness=-1}
    \label{fig: histogram of relative errors}
    \vspace{-5mm}
    % \Description{Cumulative distribution of the errors for the four methods. The curve for our library is closer to the top left corner, then GSL and lastly, Boost and std are roughly equal.}
\end{figure}

The key observation from \Cref{tab: combined_precision} is that our library never fails to compute a value, i.e., the overall robustness is $100 \%$. In addition, the median errors are at machine precision; however, the maximum relative errors are sometimes higher than the maximum relative errors for other libraries. For example, compared to Boost in the Small region for $\log I_v(x)$. Therefore, we cannot determine which library is here the most accurate. As an extension, we plot the cumulative distribution of relative errors in \Cref{fig: histogram of relative errors}, which shows that our library has a higher proportion of values with small errors. 

\Cref{tab: combined_precision} indicate some outputs from our library with high relative errors compared to the reference values of Mathematica. These are values where $v\approx 100$ and $x\approx 0.1$, and Mathematica warns about the loss of precision for such inputs. Therefore, we extracted the 35 values with the highest relative errors for our library (relative errors of $\approx 10^{-6}$ or higher), and recalculated the reference values using Wolfram$|$Alpha (\Cref{tab: precision special cases}). We see that the most accurate library is now ours, which gives answers that are 12 orders of magnitude closer to the reference values than other libraries (errors near machine precision). GSL could not calculate an answer for any of the 35 values, while std and Boost had high median and maximum errors. 

These results suggest a discrepancy between Mathematica and Wolfram$|$Alpha. We were unable to find any information about this difference, but we noted that the Wolfram 
Language supports arbitrary-precision arithmetic \citet{WolframMachinePrecision}. 
However, the Wolfram language is used by both, so this would not explain why Mathematica has numerical problems. It should be noted that Mathematica raised a warning about loss of precision due to numerical underflow, so this would support Wolfram$|$Alpha being the better option for reference values. This suggests that Mathematica may not be the optimal source for numerical tests, as is current practice \citet{Boost}.

In summary, our library has similar or better precision than existing libraries when considering the discrepancy between Mathematica and Wolfram$|$Alpha. More importantly, our library is always able to return a successful value. \looseness=-1

\begin{table*}[ht]
    \centering
    \footnotesize
    \begin{tabular}{c|c|c|c|c|c|c|c}
    \toprule
    \multirow{2}{*}{Function} & \multirow{2}{*}{Region} & \multirow{2}{*}{Metric} & \multicolumn{5}{c}{Library} \\
                            &&& std & GSL & Boost & CUSF (CPU) & CUSF (GPU) \\
                            \midrule
    \multirow{3}{*}{$\log I_v(x)$}  & \multirow{3}{*}{Selected values} & Robustness & $100 \%$ & $0 \%$ & $100 \%$ & $\bm{100 \%}$ & $\bm{100 \%}$ \\
    &&Median & $2.28 \cdot 10^{-5}$ & N/A & $2.28 \cdot 10^{-5}$ & $1.53 \cdot 10^{-16}$ & $1.53 \cdot 10^{-16}$ \\
    &&Maximum & $8.30 \cdot 10^{-4}$ & N/A & $8.30 \cdot 10^{-4}$ & $3.07 \cdot 10^{-16}$ & $3.07 \cdot 10^{-16}$ \\
    \bottomrule
    \end{tabular}
    \caption{Precision metrics for different libraries for 35 selected values in the Small region when calculating $\log I_v(x)$. 
    The robustness value in bold indicates the library with the highest robustness value, where the maximum relative error is used to break ties. The reference values are computed using Wolfram$|$Alpha. It can be seen that our library is much more precise when calculating $\log I_v(x)$ compared to the other libraries when using Wolfram$|$Alpha for the reference solutions. GSL does not return any values, and thus the median and maximum are not available (N/A).}
    \label{tab: precision special cases}
    \vspace{-2mm}
\end{table*}

Additionally, we can compare the libraries, along with the CUDA Math Library, by computing the modified Bessel function of the first kind for the special cases where $v\in\{0,1\}$; these special cases allows for specialized solutions. Our library does not have functions specifically designed to handle this special case, so the generic function for $\log I_v(x)$ is used. The metrics can be seen in \Cref{tab: precision I0} and show that std is the most precise for the values in the Small region, with our library only slightly less precise. For the Large region, the GSL and our library are the most precise, as both can compute for all input values, and the relative errors are numerical errors, while the others lack robustness and accuracy.

\begin{table*}[ht]
    \centering
    \footnotesize
    \begin{tabular}{c|c|c|c|c|c|c|c|c}
        \toprule
        \multirow{2}{*}{Function} & \multirow{2}{*}{Region} & \multirow{2}{*}{Metric} & \multicolumn{6}{c}{Library} \\
                                & & & std & GSL & Boost & CUDA & CUSF (CPU) & CUSF (GPU) \\
                                \midrule
        \multirow{6}{*}{$\log I_0(x)$}  & \multirow{3}{*}{Small}& Robustness & $\bm{100 \%}$ & $100 \%$ & $100 \%$ & $100 \%$ & $100 \%$ & $100 \%$ \\
                            && Median & 0.00 & 0.00 & 0.00 & 0.00 & 0.00 & 0.00 \\
                            && Maximum & $1.11 \cdot 10^{-14}$ & $3.08 \cdot 10^{-9}$ & $9.07 \cdot 10^{-10}$ & $9.07 \cdot 10^{-10}$ & $3.68 \cdot 10^{-13}$ & $3.68 \cdot 10^{-13}$ \\
        \cmidrule{2-9}
        &\multirow{3}{*}{Large}& Robustness & $61 \%$ & $\bm{100 \%}$ & $61 \%$ & $61 \%$ & $100 \%$ & $100 \%$ \\
                            && Median & 0.00 & 0.00 & 0.00 & 0.00 & 0.00 & 0.00 \\
                            && Maximum & $1.71 \cdot 10^{-16}$ & $2.16 \cdot 10^{-16}$ & $2.15 \cdot 10^{-16}$ & $2.15 \cdot 10^{-16}$ & $2.22 \cdot 10^{-16}$ & $2.22 \cdot 10^{-16}$ \\
        \bottomrule
    \end{tabular}
    \caption{Precision metrics for different libraries in Small and Large regions when calculating $\log I_0(x)$. The errors are the relative errors compared to the reference solutions calculated with Mathematica. The robustness value in bold indicates the library with the highest robustness value, where the maximum relative error is used to break ties. We can see that std, boost, and CUDA all lack robustness in the Large region. Results for $v=1$ are omitted as they are similar.}
    \label{tab: precision I0}
    \vspace{-5mm}
\end{table*}

\subsection{Performance}\looseness=-1
Next, we look at the runtimes of the different libraries to compare the computational efficiency of each. For the fractional-order methods, we get the runtimes shown in \Cref{tab: combined runtime for bessel functions}. From the table, we can see that our library is the fastest for computing $\log I_v(x)$ in both regions, but for the Small region, the functions from the std and GSL libraries are faster for computing $\log K_v(x)$. Except for the Small region when computing $\log K_v(x)$, the runtime of our library is about 20 to 200 ms for 10M input values. This is one to two orders of magnitude faster than the results presented by \citet{bremer2019algorithm} for computing $\log J_v(x)$ and $\log Y_v(x)$.

\begin{table*}[ht]
\centering
\begin{tabular}{l|l|c|c|c|c|c}
\toprule
\multirow{2}{*}{Function} & \multirow{2}{*}{Region} & \multicolumn{5}{c}{Library} \\
                         &                         & std & GSL & Boost & CUSF (CPU) & CUSF (GPU) \\
\midrule
\multirow{2}{*}{$\log I_v(x)$} 
                         & Small & $555.23 \pm 27.04$ & $429.50 \pm 6.83$ & $4023.40 \pm 3.44$ & $131.39 \pm 0.87$ & $\bm{50.41 \pm 0.30}$ \\
                         & Large & $26249.82 \pm 4.64$ & $73.02 \pm 6.90$ & $244023.40 \pm 35.64$ & $71.71 \pm 10.69$ & $\bm{39.68 \pm 0.22}$ \\
\midrule
\multirow{2}{*}{$\log K_v(x)$} 
                         & Small & $590.04 \pm 18.58$ & $\bm{309.61 \pm 10.75}$ & $3871.66 \pm 2.28$ & $2794.04 \pm 13.62$ & $430.60 \pm 10.60$ \\
                         & Large & $10864.81 \pm 2.72$ & $4517.38 \pm 16.03$ & $99496.82 \pm 107.38$ & $67.11 \pm 10.96$ & $\bm{39.96 \pm 0.18}$ \\
\bottomrule
\end{tabular}
\caption{Mean runtime in milliseconds over five runs for computing the modified Bessel functions of the first kind, $\log I_v(x)$, and the second kind, $\log K_v(x)$, for fractional orders in the two regions Small and Large. The $\pm$ indicates the standard deviation for the five runs. The fastest method for each function and region combination is highlighted in bold. Overall, our library is much faster than the other libraries, except for $\log K_v(x)$ in the Small region.}
\label{tab: combined runtime for bessel functions}
\vspace{-0.3cm}
\end{table*}

We can then also compare the runtimes when looking at the special case $v\in\{0,1\}$, where the CUDA Math Library is also available for a GPU comparison. For these special cases, there exist simplified expressions. The results of these tests are shown in \Cref{tab: runtime for log I0 and I1}, where it is clear that the CUDA Math Library is much faster in both regions. However, the CPU function in our library is comparable to the other CPU libraries; and for the Large region, we are much faster than the other CPU libraries.

\begin{table*}[htbp]
    \centering
    \begin{tabular}{l|l|c|c|c|c|c|c}
        \toprule
        \multirow{2}{*}{Function} & \multirow{2}{*}{Region} & \multicolumn{6}{c}{Library} \\
         &       & std & GSL & Boost & CUDA & CUSF (CPU) & CUSF (GPU) \\\midrule
   \multirow{2}{*}{$\log I_0(x)$} & Small & $3512.69 \pm 6.57$ & $915.87 \pm 10.33$ & $903.93 \pm 26.38$ & $\bm{61.56 \pm 1.88}$ & $1665.29 \pm 5.96$ & $433.00 \pm 18.63$ \\
    & Large & $19560.23 \pm 14.13$ & $929.41 \pm 24.00$ & $2080.96 \pm 33.07$ & $\bm{50.15 \pm 0.26}$ & $301.89 \pm 25.13$ & $187.14 \pm 21.57$ \\
        \midrule
    \multirow{2}{*}{$\log I_1(x)$} & Small & $3480.81 \pm 4.66$ & $966.92 \pm 77.46$ & $823.42 \pm 7.63$ & $\bm{44.43 \pm 0.03}$ & $1285.75 \pm 16.32$ & $336.44 \pm 2.79$ \\
    & Large & $19620.46 \pm 53.96$ & $946.19 \pm 23.20$ & $1937.93 \pm 7.65$ & $\bm{36.28 \pm 0.77}$ & $261.30 \pm 15.74$ & $175.59 \pm 0.51$ \\
        \bottomrule
    \end{tabular}
    \caption{Mean runtime in milliseconds over five runs when computing $\log I_0(x)$ and $\log I_1(x)$. The $\pm$ indicates the standard deviation for the five runs. The GNU Scientific Library (GSL), Boost, and CUDA use special functions for $\log I_0(x)$ and $\log I_1(x)$. Our library uses the general $\log I_v(x)$. The fastest library is CUDA, and our GPU version is second. Our CPU version is comparable to the other CPU libraries. We can see that there is only a small difference between $v=0$ and $v=1$.}
    \label{tab: runtime for log I0 and I1}
    \vspace{-0.3cm}
\end{table*}

\subsection{Metric learning}\label{sec: metric learning}

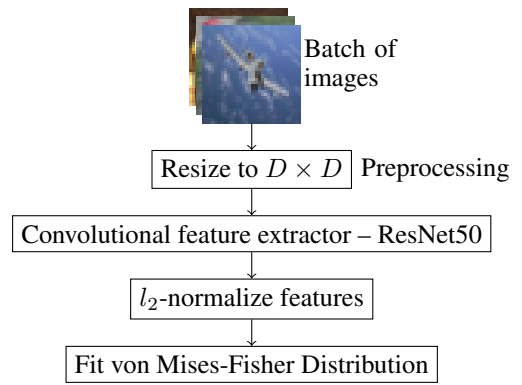
\begin{figure}[ht]
\centering
\begin{tikzpicture}[node distance=0.35cm, auto]
    % Define style for images
    \tikzstyle{image} = [anchor=south west,inner sep=0]

    % Define nodes
    
    \node [image] (image1) {\includegraphics[width=1.3cm]{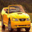}};
    \node [image, below left=-1.2cm and -1.40cm of image1] (image2) {\includegraphics[width=1.3cm]{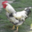}};
    \node [image, below left=-1.2cm and -1.40cm of image2] (image3) {\includegraphics[width=1.3cm]{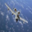}};
    \node[anchor=west, text width=1.7cm] at (image2.east) {Batch of images};

    \node [draw, below=of image3, align=center] (resize) {Resize to $D \times D$};
    \node[anchor=west] at (resize.east) {Preprocessing};
    \node [draw, below=of resize, align=center] (resnet) {Convolutional feature extractor -- ResNet50};
    \node [draw, below=of resnet, align=center] (normalize) {$l_2$-normalize features};
    \node [draw, below=of normalize, align=center] (fitting) {Fit von Mises-Fisher Distribution};

    % Connect nodes
    \draw [->] (image3.south) -- (resize.north);
    \draw [->] (resize) -- (resnet);
    \draw [->] (resnet) -- (normalize);
    \draw [->] (normalize) -- (fitting);
\end{tikzpicture}
\caption{Metric learning pipeline. Images from CIFAR10 ( \Cref{fig:cifar10 examples}) are resized to three different sizes of $D=32$, $D=64$, and $D=128$, and passed through the convolutional layers from ResNet50 to extract image features. We fit a vMF distribution to the $l_2$-normalized features. The extracted features have 2048, 8192, and 32768 dimensions, respectively.}
% \Description{Flow chart of the data processing, showing that a batch of images are taken in, then they are resized to size D by D, features are extracted using ResNet50, the features are normalized to lie on the unit hypersphere, and lastly the images are used to fit vMF distributions. }
\vspace{-0.5cm}
\label{fig:data_pipeline}
\end{figure}

We now want to show that our library can be used to solve a real-world problem using a setup similar to \citet{warburg:metric:2023}. Here, we fit high-dimensional features from image processing to von Mises Fischer, vMF, distributions. We choose this distribution because it models data that exist on a unit hypersphere in $p$-dimensional space, $S^{p-1}=\{ x\in\mathbb{R}^p | \|x\|_2 = 1 \}$. This is exactly the type of data that results from $l_2$-normalization of the extracted features \citet{musgrave2020metric}. 

% Pipeline
% We process the images by resizing them to a resolution of $D \times D$, and then apply the convolutional layers from an image classification model. This yields high-dimensional features, and we can control the feature dimension by changing the resolution $D$ of the images due to the convolutional layers. 
The images are first resized to a resolution of $D \times D$. We then use the convolutional layers of a pre-trained image classification model to extract high-dimensional features. This pipeline has been shown in \Cref{fig:data_pipeline}. The dimensionality of the extracted features is adjustable by changing the resolution $D$.

\begin{figure}
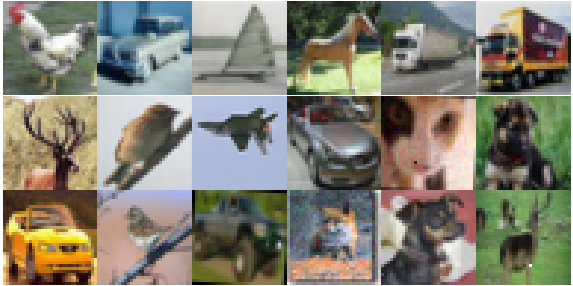

    \centering
    \include{Figures/cifar10_example}
    \vspace{-8mm}
    \caption{Example images from the CIFAR10 dataset.}
    \label{fig:cifar10 examples}
    % \Description{Example of 18 images from the CIFAR10 dataset in an 3 by 6 grid. The images show a chicken, car, boat, toy horse, truck, truck, deer with antlers, bird, fighter jet, car, cat, dog, car, bird, car, fox, dog, and deer.}
    \vspace{-0.4cm}
\end{figure}

% Data
We use the CIFAR10 training dataset of $32 \times 32$ color images, which consists of 50k images in ten categories: airplane, automobile, bird, cat, deer, dog, frog, horse, ship, and truck (see \citet{krizhevsky2009learning} for details). Example images are shown in \Cref{fig:cifar10 examples}. The images are resized to affect the dimensionality of the extracted features to which we fit a vMF distribution. We bilinearly resize the images to $32 \times $, $64 \times 64$, and $128 \times 128$.
% Image classification model
To extract features from the images, we use a convolutional model as it is agnostic to the input dimensions and gives output dimensions based on the input dimensions. For this project, we treat the convolutional model as a black-box feature extractor. We used the convolutional layers from the ResNet50 model with pre-trained weights, as this is a widely used baseline computer vision model \citet{he2016deep, pytorch_blog}. Given the sizes of the images used, the extracted features have sizes 2048, 8192, and 32768. 

% Information on the vMF distribution.
As mentioned earlier, the von Mises-Fischer (vMF) distribution models data on the sphere, where it generalizes the normal distribution \citet{mardia2000directional, gatto2011generalized}. 
The distribution is given by a mean direction, $\bm{\mu}$, and concentration parameter, $\kappa$, and has density
\begin{align*}
    f_p(\mathbf{x} | \boldsymbol{\mu}, \kappa) &= C_p(\kappa) \exp{(\kappa \boldsymbol{\mu}^\top \mathbf{x})},\ %\label{eq:pdf vMF} \\
    % f_p(\bm{x}; \bm{\mu},\kappa) &= C_p(\kappa) \exp(\kappa \bm{\mu}^T\bm{x}) \\
    C_p(\kappa) = \frac{\kappa^{p/2-1}}{(2\pi)^{\frac{p}{2}} I_{p/2-1}(\kappa)}.
\end{align*}
We see above that we need $I_v(x)$ for $v=\sfrac{p}{2}-1$ to compute the probability density function for the vMF distribution. There are explicit expressions to fit the parameters $\bm{\mu}$ and $\kappa$ to a dataset by maximizing the log-likelihood. The mean direction is given by taking the mean of the data. Suppose that we have the data $X=\{ x_i|x_i\in\mathbb{R}^p,\|x\|_2 = 1 \}$, then we can estimate $\bm{\mu}$,\looseness=-1
\begin{align}
    \bm{\mu} = \frac{\overline{x}}{\overline{R}}, \quad \overline{x}=\frac{1}{|X|}\sum_{x_i\in X} x_i, \quad \overline{R} = \|\overline{x}\|_2.\label{eq: mean direction estimate}
\end{align}
The concentration parameter $\kappa$ is more difficult to estimate. However, \citet{sra2012short} gave an approximation, $\widehat{\kappa_0}$, which can be further improved by performing one or two Newton updates, $\widehat{\kappa_1}$ and $\widehat{\kappa_2}$, respectively; these are given below.
% \begin{align}
%     \widehat{\kappa_0} &= \frac{\overline{R}(p - \overline{R}^2)}{1 - \overline{R}^2}, \label{eq: kappa0 estimate}\\
%     \widehat{\kappa_1} &= \widehat{\kappa_0} - \frac{A_p(\widehat{\kappa_0}) - \overline{R}}{1 - A_p(\widehat{\kappa_0})^2 - \frac{p-1}{\widehat{\kappa_0}}A_p(\widehat{\kappa_0})}, \label{eq: kappa1 estimate}\\
%     \widehat{\kappa_2} &= \widehat{\kappa_1} - \frac{A_p(\widehat{\kappa_1}) - \overline{R}}{1 - A_p(\widehat{\kappa_1})^2 - \frac{p-1}{\widehat{\kappa_1}}A_p(\widehat{\kappa_1})}, \label{eq: kappa2 estimate}\\
%     A_p(\widehat{\kappa}) &= \frac{I_{p/2}(\widehat{\kappa})}{I_{p/2-1}(\widehat{\kappa})}.\nonumber
% \end{align}
\begin{align}
    F(\kappa) &= \kappa - \frac{A_p(\kappa) - \overline{R}}{1 - A_p(\kappa)^2 - \frac{p-1}{\kappa}A_p(\kappa)},
    A_p(\widehat{\kappa}) = \frac{I_{p/2}(\widehat{\kappa})}{I_{p/2-1}(\widehat{\kappa})},\nonumber\\
    \widehat{\kappa_0} &= \frac{\overline{R}(p - \overline{R}^2)}{1 - \overline{R}^2},\quad
    \widehat{\kappa_1} = F(\widehat{\kappa_0}),\quad
    \widehat{\kappa_2} = F(\widehat{\kappa_1}). \label{eq: kappa estimates}
\end{align}
% In their paper, the author performs numerical experiments to assess the accuracy of the estimates. They find the relative error of $\widehat{\kappa_2}$ to be around $10^{-12}$ to $10^{-11}$ across dimensions $p\in[100,100000]$ \cite[Table 2]{sra2012short}.
Numerical evaluations in their paper show a relative error for $\widehat{\kappa_2}$ between $10^{-12}$ and $10^{-11}$ over dimensions $p\in[100,100000]$ \cite[Table 2]{sra2012short}.

We focus on the feasibility of fitting the vMF distributions to evaluate our library. We fit the mean direction parameter using \Cref{eq: mean direction estimate} to the data from the data pipeline. We then maximize the log-likelihood over the concentration parameter $\kappa$ given the mean direction and compare the resulting estimate to the approximations $\widehat{\kappa_i},i=0,1,2$. The log-likelihood is given by,\looseness=-1
\begin{align*}
    \text{logLik}(\kappa|X) = \frac{1}{|X|} \sum_{x_i\in X}& \log f_p(\mathbf{x} | \boldsymbol{\mu}, \kappa) \\
    = \frac{1}{|X|}\sum_{x_i\in X}&\left[ \left(\frac{p}{2}-1\right)\log\kappa - \frac{p}{2}\log 2\pi\right. \\
    &-\left. \log I_{p/2-1}(\kappa) + \kappa \boldsymbol{\mu}^\top \mathbf{x} \right].
\end{align*}

We maximize the log-likelihood by minimizing the negative log-likelihood using \texttt{minimize} in SciPy with the L-BFGS-B method, bounding $\kappa$ to be a non-negative number. The optimization is performed with and without analytic gradients. The results of the optimizations can be seen in \Cref{table:vmf-results}. For the gradient-free method, we see that the estimated $\kappa$ matches the first six, four, and two digits of $\widehat{\kappa_2}$ for the 2048-, 8192-, and 32768-dimensional features, respectively. The optimizer stops because the gradient is estimated to be zero, which is caused by numerical errors, since the negative log-likelihood is still decreasing, with the minimum close to $\widehat{\kappa_2}$. However, even with these numerical inaccuracies, the estimates are still within $0.2 \%$ of the best available estimate. With the gradient, we see that the optimizer can estimate $\kappa$ much better, with relative errors $\leq 3.87\cdot 10^{-11}$. Thus, with the gradient, the relative errors match the errors reported by \citet{sra2012short} for their method, so we cannot rule out that our errors are caused by inaccuracies in $\widehat{\kappa_2}$. 

We performed the same test using SciPy and mpmath to compute the Bessel functions; however, the optimizer could not converge in any tests. With mpmath, it would abort due to loss of precision, and with SciPy it would diverge. Thus, only with our library can we estimate the parameters of a vMF distribution by optimizing the log-likelihood to a dataset. This demonstrates that our approach opens up tasks involving the modified Bessel functions of the first and second kind that were previously not realistic. 

\begin{table}
    \centering
    \begin{tabular}{c|c c c }
    \toprule
    \# of features & 2048 & 8192 & 32768 \\\midrule
    Gradient free estimate  & 298.9091 & 1577.135 & 6681.31 \\\hline
    Gradient estimate       & 298.9098 & 1577.405 & 6668.07 \\\midrule
    $\widehat{\kappa_0}$    & 298.9127 & 1577.412 & 6668.08 \\\hline
    $\widehat{\kappa_1}$    & 298.9098 & 1577.405 & 6668.07 \\\hline
    $\widehat{\kappa_2}$    & 298.9098 & 1577.405 & 6668.07 \\\bottomrule
    \end{tabular}
    % \vspace{0.1cm}
    \caption{Results for fitting the vMF distribution to assess our implementation of $\log I_v(x)$ in a real-world use case. The table presents estimates obtained by fitting the mean direction parameter according to \Cref{eq: mean direction estimate} and maximizing the log-likelihood over the concentration parameter $\kappa$. The estimates are compared with the approximations $\widehat{\kappa_i}, i=0,1,2$ from \Cref{eq: kappa estimates}.% \Crefrange{eq: kappa0 estimate}{eq: kappa2 estimate}. 
    The gradient-free estimates have a relative error of $2.39\cdot10^{-6}$, $1.71\cdot10^{-4}$ and $1.99\cdot10^{-3}$ for, respectively, \mbox{2048-,} 8192- and 32768-dimensional features compared to $\widehat{\kappa_2}$. With gradient, the relative errors drop to, respectively, $3.87 \times 10^{-11}$, $2.13 \times 10^{-11}$, $1.72 \times 10^{-11}$.\looseness=-1}
    \label{table:vmf-results}
    \vspace{-.8cm}
\end{table}

\section{Conclusion}

This paper presents new algorithms for computing the logarithm of modified Bessel functions of the first and second kind that address critical precision limitations and underflow and overflow problems in existing libraries that make existing libraries unsuitable for many practical applications. 
Our algorithms consistently produced numerically stable results without underflow or overflow, with precision equal to or better than current C++ libraries and significantly faster running times. In most cases, the runtime was one or two orders of magnitude faster.
These results have implications for the use of the modified Bessel functions in practical applications. 
We present an example use case by successfully fitting the von Mises-Fisher distribution to high-dimensional data by numerically optimizing the log-likelihood of the data. 
This demonstrates that our libraries enable the use of modified Bessel functions for high-dimensional data, which was previously not possible with available libraries.\looseness=-1

Our methods significantly outperform existing libraries that compute exponentially scaled functions, especially in terms of robustness and computational efficiency. 
The only exception is the modified Bessel function of the second kind for small values, where our library currently lags behind existing solutions in speed and accuracy. However, our library is still more robust than existing libraries.
Future research could explore the potential of developing functions that are specialized for certain inputs, such as integer order. 
In addition, implementing derivatives of these functions would facilitate the use of gradient-based optimization techniques, further extending the utility of our library. 
There is also potential to adapt our approach to other specialized functions that face similar computational challenges.

\textbf{Acknowledgements.~~}
This work was supported by a research grant (42062) from VILLUM FONDEN. This project received funding from the European Research Council (ERC) under the European Union’s Horizon 2020 research and innovation programme (grant agreement 757360). The work was partly funded by the Novo Nordisk Foundation through the Center for Basic Machine Learning Research in Life Science (NNF20OC0062606).

\bibliography{main}

\end{document}